\documentclass[onecolumn]{emulateapj}
\slugcomment{Astro-ph archives (submitted: 30$^{th}$ Jul 2014)}
\shorttitle{IMPULSIVE ENERGY RELEASE IN AN M4.0 CONFINED FLARE}
\shortauthors{KUSHWAHA ET AL.}
\begin{document}
\title{Impulsive energy release and non-thermal emission in a confined M4.0 flare triggered by rapidly evolving magnetic structures}
\author{Upendra Kushwaha and Bhuwan Joshi}
\affil{Udaipur Solar Observatory, Physical Research Laboratory, Udaipur 313 001, India}
\author{Kyung-Suk Cho}
\affil{Korea Astronomy and Space Science Institute, Daejeon 305-348, Korea}
\author{Astrid M. Veronig}
\affil{Kanzelh\"ohe Observatory/Institute of Physics, University of Graz, Universit$\ddot{a}$tsplatz 5, A-8010 Graz, Austria}
\author{Sanjiv Kumar Tiwari}
\affil{NASA Marshall Space Flight Center, ZP 13, Huntsville, AL, 35812, USA}

\author{S.K. Mathew}
\affil{Udaipur Solar Observatory, Physical Research Laboratory, Udaipur 313 001, India}

\email{upendra@prl.res.in}
 
 \begin{abstract}
 
We present observations of a confined M4.0 flare from NOAA 11302 on 2011 September 26. Observations at high temporal, spatial and spectral resolution from Solar Dynamics Observatory, Reuven Ramaty High Energy Solar Spectroscopic Imager, and Nobeyama Radioheliograph enabled us to explore the possible triggering and energy release processes of this flare despite its very impulsive behavior and compact morphology. The flare light curves exhibit an abrupt rise of non-thermal emission with co-temporal hard X-ray (HXR) and microwave (MW) bursts that peaked instantly without any precursor emission. This stage was associated with HXR emission up to 200 keV that followed a power law with photon spectral index ($ \gamma)$ $\sim $3. Another non-thermal peak, observed 32 s later, was more pronounced in the MW flux than the HXR profiles. Dual peaked structure in the MW and HXR light curves suggest a two-step magnetic reconnection process. Extreme ultraviolet (EUV) images exhibit a sequential evolution of the inner and outer core regions of magnetic loop system while the overlying loop configuration remained unaltered. Combined observations in HXR, (E)UV, and H$ _{\alpha} $ provide support for flare models involving interaction of coronal loops. The magnetograms obtained from Helioseismic and Magnetic Imager (HMI) reveal the emergence of magnetic flux which started $ \sim $5 hr before the flare. However, the more crucial changes in the photospheric magnetic flux occurred about 1 min prior to the flare onset with opposite polarity magnetic transients appearing at the early flare location within the inner core region. The spectral, temporal, and spatial properties of magnetic transients suggest that the sudden changes in the small-scale magnetic field have likely triggered the flare by destabilizing the highly sheared pre-flare magnetic configuration.
    
\end{abstract}


\section{Introduction}

Solar flares are the most striking explosive form of solar activity. A flare is characterized by a catastrophic release of huge amounts of energy (up to  $\sim$10$ ^{32}$ ergs) on timescales of a few seconds to tens of minutes. Flares manifest their signatures in a wide range of electromagnetic spectrum, from radio to $\gamma$-rays, and involve substantial mass motions and particle acceleration \citep[for a review see][]{Fletcher2011}. It is now well established that the energy released during flares is stored in the corona prior to the event in the form of stressed or non-potential magnetic fields. Magnetic reconnection has been recognized as the fundamental process responsible for the changes in the topology of magnetic fields, as well as the rapid conversion of stored magnetic energy into thermal and kinetic energy of plasma and particles during a flare \citep[e.g., see reviews by][]{Priest2002,LinJ2003}. Investigations of the causes and consequences of solar flares are of immense importance to broaden our understanding of the space weather.

The very early observations of solar flares in soft X-ray (SXR) wavelengths from
the $Skylab$ mission in 1973--1974 established two morphologically distinct classes of flares: confined and eruptive events \citep{Pallavicini1977}. The confined flares show brightening in compact loop structures that lasts for only a short period (e.g., on the order of a few minutes). They are generally modeled in terms of energy release within a single static magnetic loop and are thus also referred to as single-loop or compact flares. The second category comprises the long-duration events (LDE) in SXRs (e.g., tens of minutes up to hours) which are eruptive in nature. Motions of large chromospheric flare ribbons and the formation of an arcade of (post-flare) loops during LDE flares essentially reveal large-scale restructuring of coronal loops. After the advent of direct coronal mass ejection (CME) observations, it has been recognized that LDE flares are strongly associated with CMEs while confined events lack CMEs. It is worth mentioning that the temporal evolution of CME and flare signatures in eruptive events suggests that both phenomena have a strongly coupled relationship but not a cause-effect one \citep{Zhang2001,Temmer2010}. In fact there are instances when highly energetic X-class flares occur without associated CMEs \citep{Wang2007}. 

It has been commonly observed that solar flares occur in closed magnetic field configurations associated with active regions. It is very likely that such closed magnetic structures will encompass one or more magnetic neutral lines in the photospheric magnetic flux. In a simplistic model, one can think of a bipolar magnetic configuration in terms of an inner region, called the core fields, and the outer region, called the envelope or overlying fields \citep{Moore2001}. The core fields are rooted close to the neutral line while the envelope fields are rooted away from it. The core fields are usually strongly non-potential in the pre-flare phase. The initial energy release occurs in core region while the overlying fields act as a constraining force to prevent the eruption of core fields \citep[see, e.g.,][]{Wang2007}. 

The basic magnetic configuration during a solar flare is revealed by magnetic loops which can be quite complex in some cases. Observations also indicate that in some cases magnetic reconnection takes place between pre-existing loops and newly emerging flux \citep{Hanaoka1997}. The multi-wavelength investigations of the evolution of coronal loops and changes in their connectivities during a flare provide important constraints on the role of magnetic reconnection during the restructuring of the magnetic configuration \citep[e.g.,][]{Hanaoka1996,Kundu2001,Sui2006,Su2013}. Understanding the physical conditions of the core region of a flare's magnetic environment can provide special insights into the underlying mechanism for impulsive energy release \citep[see e.g.,][]{Chandra2009}. Flare research has been dominated by the study of eruptive flares because of their large-scale structure and long duration. On the other hand, it is rather challenging to investigate energy release processes in confined flares due to their rapid evolution in a compact region that imposes severe observational constraints. Now with availability of solar observations from Solar Dynamics Observatory ($SDO$) and Reuven Ramaty High Energy Solar Spectroscopic Imager  ($RHESSI$) having unprecedented observational capabilities in temporal, spatial, and spectral domains, it would be quite promising to perform an in depth analysis of energetic confined flares despite their compact and abrupt nature.
 
Decades-long observations have established that the fast evolution of an active region is associated with  the occurrence of flares/CMEs \citep[for a review see][]{vanDriel-Gesztelyi2009}. The emergence of new flux through the photosphere plays a crucial role in the rapid evolution of active region. During majority of large flares, significant changes in the photospheric magnetic fields have been observed in large flares to occur in large flares within 10 minutes of flare onset \citep{Sudol2005}. \citet{Wang2002} reported a very striking evidence for changes in photospheric magnetic fields before the flare onset in the form of rapid disappearance of a small sunspot at the flaring site. Recently, \cite{Kusano2012} have systematically shown that the evolution of small-scale magnetic patches can disturb the coronal magnetic field through reconnection and produce a large-scale flare. Now with the availability of high time cadence longitudinal magnetograms from the Helioseismic Magnetic Imager (HMI) on board SDO, it has became possible to precisely examine such magnetic fields changes before and during flares. 
     
In this paper, we  perform a detailed multi-wavelength study of a confined M4.0 flare that occurred on 2011 September 26. This flare was observed by $SDO, RHESSI$ and Nobeyama Radioheliograph (NoRH). Although the event was short-lived and confined, it produced intense non-thermal emission in hard X-ray (HXR) at energies up to 200 keV. The photospheric line of sight (LOS) magnetic field measurements from HMI reveal significant variations in the magnetic structure of the activity site before and during the event. In Section~\ref{sec:obs_analysis}, we present the instruments and data used in this study. In Section~\ref{analysis_result}, we analyze multi-wavelength data which essentially emphasize the link between various flare-associated phenomena from pre- to post-flare stages in different layers of the solar atmosphere. The discussion and conclusions are presented in Section~\ref{sec:res}.

\begin{figure}
\epsscale{1.0}
\plotone{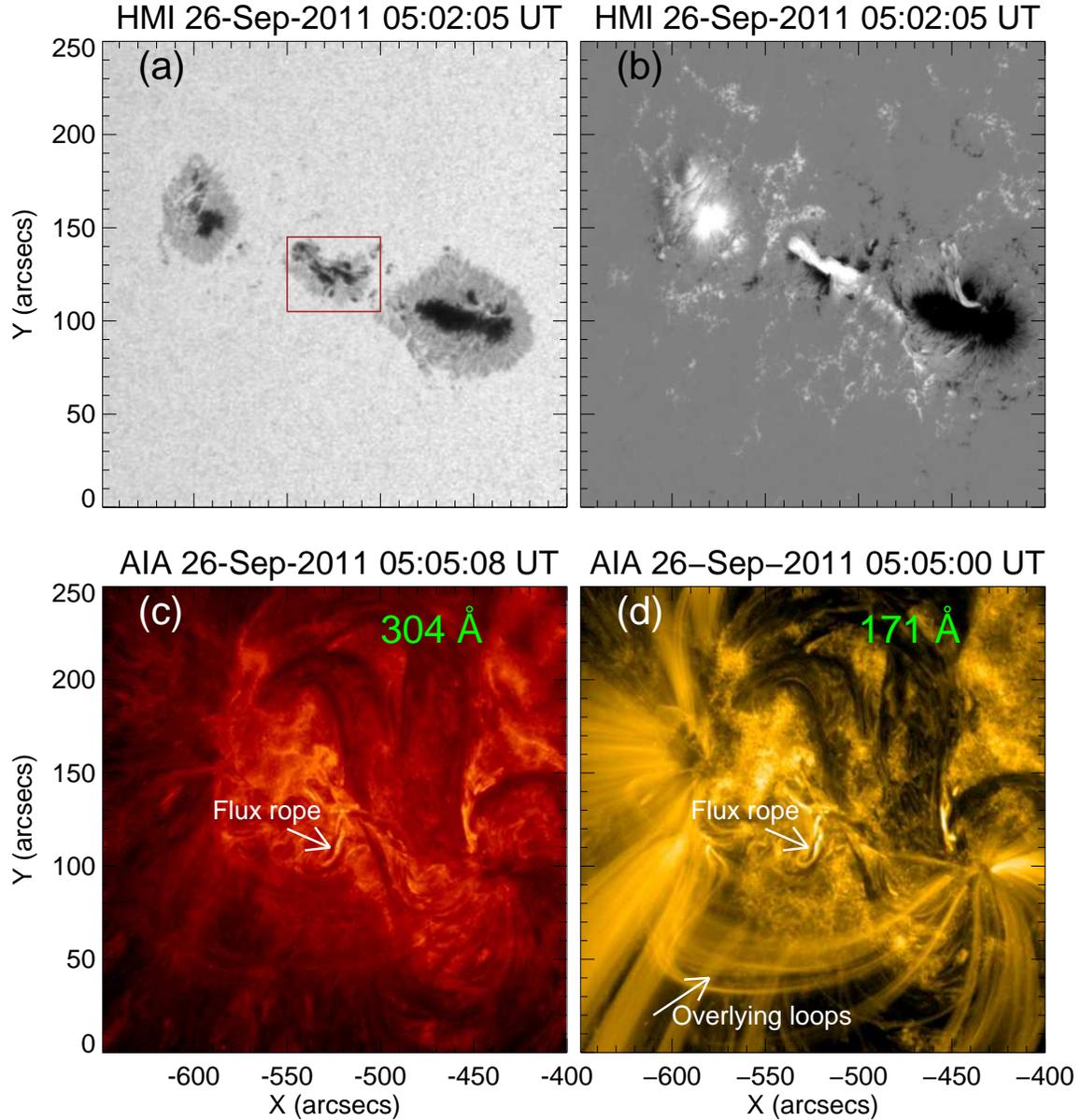}
\caption{Multi-wavelength view of the active region NOAA 11302 on 2011 September 26. Panel (a): HMI white light picture of the active region presenting its elongated shape with three well-developed sunspot groups. The M4.0 flare is associated with the central sunspot group shown by a rectangular box. Panel (b): HMI magnetogram of the active region showing that the leading and following sunspot groups have negative and positive polarity magnetic flux, respectively. However, in between these two sunspot groups a small sunspot group has a complex mix polarity. Panel (c): AIA 304 \AA~ image of NOAA 11302 just before the flare onset. A small flux rope is noted at the source region and is indicated by an arrow. Panel (d): AIA 171 \AA~image clearly shows that the source region was fully enveloped the overlying loop system that is marked by another arrow.}
\label{event_overview}
\end{figure}

\section{OBSERVATIONS}
\label{sec:obs_analysis}

Our analysis is based on the observations taken by the following set of instruments: the Atmospheric Imaging Assembly (AIA, \citealt{Lemen2012}) and HMI \citep{Schou2012} on board $SDO, RHESSI$ \citep{Lin2002}, and NoRH \citep{Nakajima1994a,Takano1997}. AIA records full-disk images at 12 s cadence in seven different EUV filters (94, 131, 171, 193, 211, 304, and 335~\AA), at 24 s cadence in two UV filters (1600 and 1700 \AA), and at 3600 s cadence in a white light filter (4500 \AA). In this paper, we present AIA images taken in 171, 304, 131, 94, 1600 and 1700 \AA~band passes with a spatial sampling of 0$^{\prime\prime}$.6~pixel$^{-1}$. During the flare activity, AIA takes images with different exposure times ranging from 0.1 to 2.9 s. To compensate for this, each AIA image was normalized to its respective exposure time. The magnetic structure and evolution of the active region was studied with HMI LOS and vector magnetograms. HMI observes the full solar disk in the photospheric absorption line centered at 6173.3~ \AA~(Fe {\small I}) and produces images with a spatial and temporal resolutions of $0^{\prime\prime}$.5 pixel$^{-1}$ and 45 s, respectively.

$RHESSI$ observes the full Sun with an unprecedented combination of spatial resolution (as fine as $ \sim $2$^{\prime\prime}$.3) and energy resolution (1--5 keV) in the energy range of 3 keV to 17 MeV. To reconstruct X-ray images, we have used the computationally expensive PIXON algorithm \citep{Metcalf1996}, which is thought to provide the most accurate image photometry \citep{Alexander1997}. The images are reconstructed by selecting front detector segments 3--8 (excluding 7) with 20 s integration time. The flare is also completely observed in microwave (MW) frequencies by NoRH at 17 GHz and 34 GHz. At these two frequencies NoRH has spatial resolutions of 10$^{\prime\prime} $ and 5$^{\prime\prime} $, respectively. NoRH has a sensitivity of at least 1 solar flux unit (sfu) at 17 GHz and $\sim $3 sfu at 34 GHz. The normal time resolution of NoRH is 1 s, but a resolution as good as 50 ms can be used for special projects \citep{Kundu2006}. For the present study, we have analyzed NoRH light curves and images with a temporal cadence of 1 s.
 
 \begin{figure}
\epsscale{1.0}
\plotone{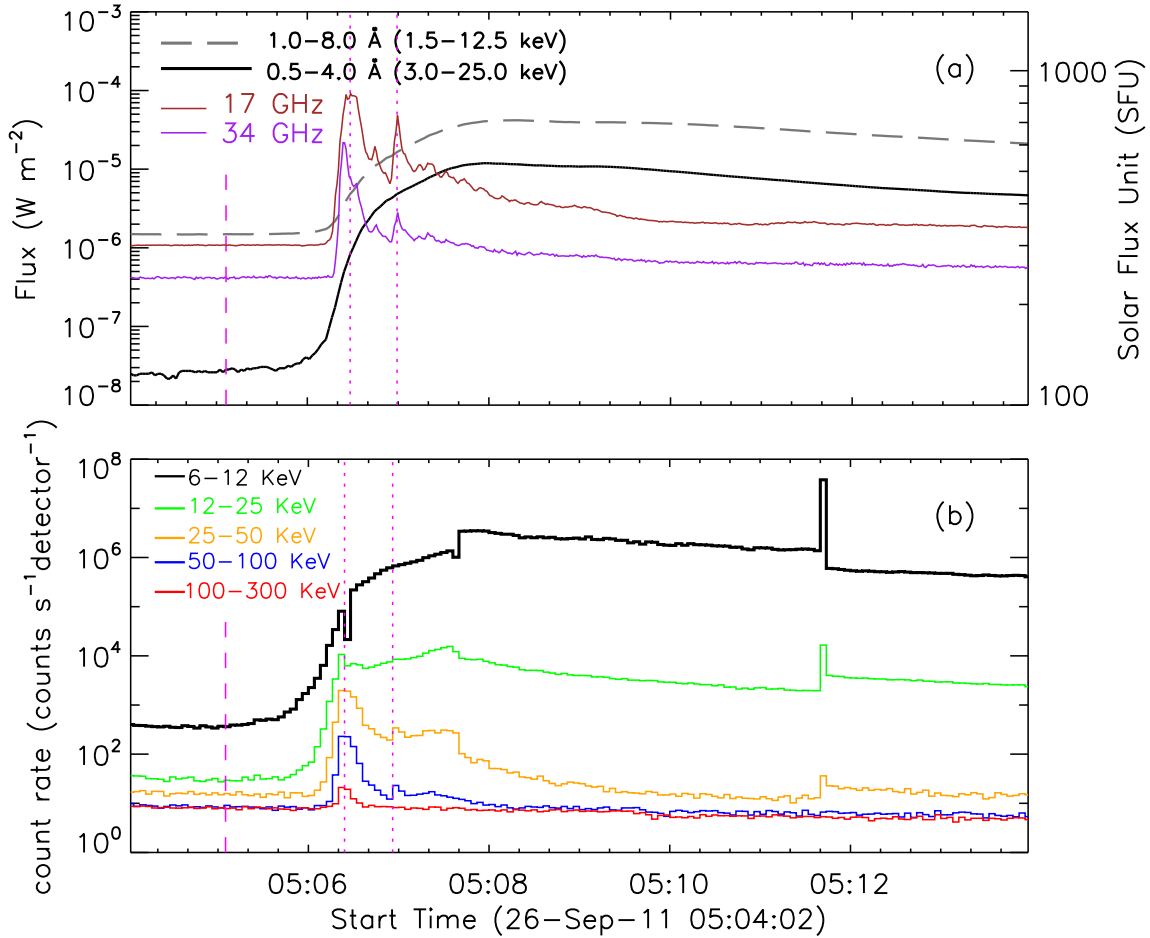}
\caption{$GOES$ soft X-ray flux profile, overlaid by NoRH microwave fluxes (top panel) and $RHESSI$ X-ray light curves (bottom panel) of an M4.0 class flare on 2011 September 26. In order to present $RHESSI$ light curves with clarity in different energy bands, count rates are scaled by factors of 1, 1/2, 1/5, 1/10, and 1/40 for the energy bands 6--12 keV (black), 12--25 keV (green), 25--50 keV (orange), 50--100 keV (blue), and 100--300 keV (red), respectively. The dashed vertical line marks the time at which flux cancellation begins in sub-region S1 (see Figure \ref{fig_hmi_bmts} (d)). Another two dotted vertical lines indicate consistent peaks in the HXR and MW light curves. The spikes seen in $RHESSI$ light curves at 05:07:50 UT and  05:11:20 UT correspond to instrumental artifact due to the change in attenuator state.}
\label{rhessi_goes_lc_2}
\end{figure}

\section {DATA ANALYSIS AND RESULTS}
\label{analysis_result} 
 \subsection{Event overview}
 \label{overview}
According to the solar region summary reports compiled by the Space Weather Prediction Center\footnote{http://www.swpc.noaa.gov}, an unnumbered active region was encroaching over the eastern limb on 2011 September 21 with an uncertain magnetic configuration. It was designated as NOAA 11302 with $\beta\gamma$ magnetic configuration on 2011 September 23. The active region developed from $\beta\gamma$ to a complex $\beta\gamma\delta$ configuration during September 25--26. NOAA 11302 was indeed a highly flare-productive active region which produced 2 X-class and 14 M-class flares within a week (i.e., during the period 2011 September 22--28). The impulsive M4.0 flare, analyzed in this paper, occurred between 05:06 UT and 05:14 UT on 2011 September 26 in this region when the average position of the active region on the solar disk was N13E34.

In Figure \ref{event_overview}, we present the multi-wavelength view of active region NOAA 11302. The active region comprised of three well-developed sunspot groups as shown in HMI continuum image (see Figure \ref{event_overview}(a)). The flaring site is enclosed by a rectangular box. The HMI magnetogram reveals that the leading sunspot group had a strong negative polarity, while the following sunspot group was dominated by strong positive polarity. However, the flare occurred in a mixed polarity region which lies in between the leading and trailing sunspot groups (see Figure \ref{event_overview}(a) and (b)). Figures~\ref{event_overview}(c) and (d) present images of the active region 
in the 304 \AA~and 171 \AA~channels of AIA, respectively. We note a small flux rope in EUV images at the flare location which is marked by arrows in Figure \ref{event_overview}(c) and (d). The observations in AIA 171 \AA~clearly indicate that the flaring site was enveloped by overlying coronal loops which are shown by an arrow (see Figure \ref{event_overview}(d)). 

\begin{figure}
\plotone{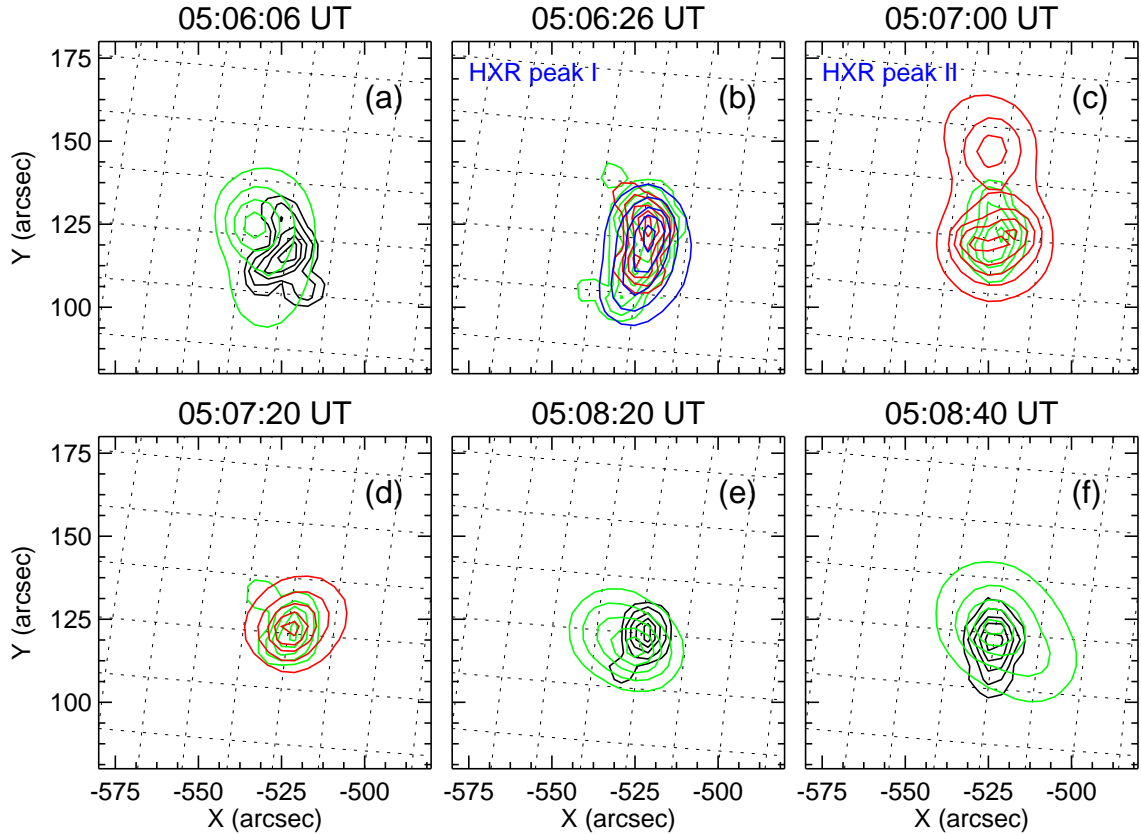}
\caption{Evolution of HXR sources in 12--25 keV (black), 25--50 keV (green), 50--100 keV (red), and 100--200 keV (blue) energy bands. $RHESSI$ images are reconstructed with the PIXON algorithm with an integration time of 20 s (mid-time is mentioned at the top of each panel). The contour levels for the images are set as 15$\%$, 30$\%$, 50$\%$, 70$\%$, and 90$\%$ of the peak flux for each image.}
\label{X-ray_mosaic}
\end{figure}

\subsection{X-ray and MW observations}
\label{sub:XR-MW}

In Figure \ref{rhessi_goes_lc_2}, we present X-ray and MW light curves of the flare in different energy bands. Figure~\ref{rhessi_goes_lc_2}(a) shows $GOES$ X-ray time profiles (in 1--8~\AA~and 0.5--4.0~\AA~energy bands) and NoRH MW light curves (in 17 GHz and 34 GHz frequencies). According to $GOES$ reports, the flare took place between 05:06 UT and 05:14 UT with a peak at $\sim $05:08 UT. We note that MW emissions at both frequencies (i.e., 17 GHz and 34 GHz) exhibit two distinct peaks at an interval of $ \sim $32 s. The $RHESSI$ light curves in five different energy bands are shown in Figure \ref{rhessi_goes_lc_2}(b). The HXR emissions ($ >$25~keV) show a prominent peak at 05:06:24 UT (peak I) which is well consistent with the first (strongest) MW bursts.  Furthermore, we note that HXR light curves in the 25--50 and 50--100 keV energy bands exhibit a second peak (peak II), again showing consistency with the second MW bursts at 05:06:56 UT. These two consistent peaks in HXR and MW are marked by the dotted vertical lines (see Figure \ref{rhessi_goes_lc_2}). On the other hand, the HXR emissions at lower energies (i.e., $ < $25 keV) resemble $GOES$ SXR profiles.

\begin{figure}
\epsscale{0.90}
\plotone{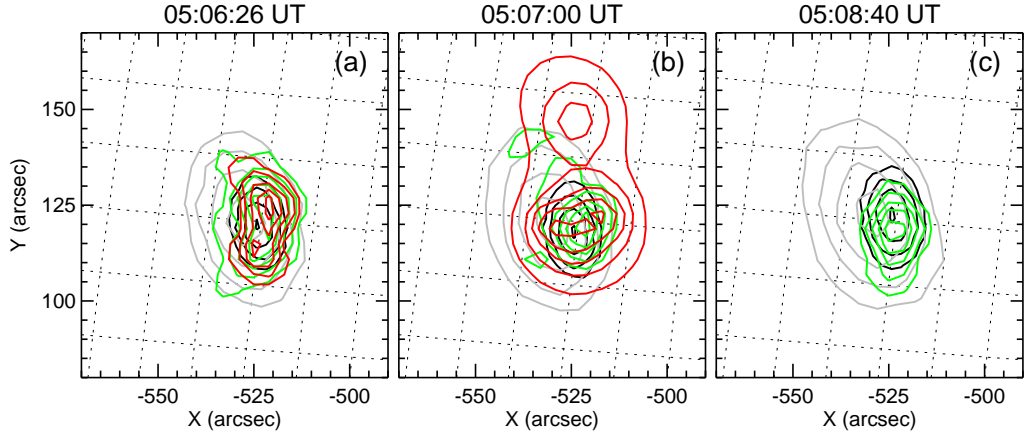}
\caption{Spatial evolution of HXR sources (12--25 keV: green, 50--100 keV: red) with respect to the MW emission (17 GHz: gray, 34 GHz: black). The contour levels for $RHESSI$ PIXON images are 15 $ \% $, 30$\%$, 50$\%$,  70$\%$, and 90$\%$ of peak flux for each image. For NoRH MW images, contour levels are set as 15$\%$, 30$\%$, 50$\%$, and 85$\%$ of the peak flux.}
\label{X-ray_MW_images}
\end{figure}

In Figure~\ref{X-ray_mosaic}, we present a sequence of $RHESSI$ X-ray images that show the spatial evolution of HXR source in different energy bands, viz., 12--25~keV (black), 25--50~keV (green), 50--100~keV (red), and 100--200~keV (blue). The $RHESSI$ images are reconstructed using the PIXON algorithm with the natural weighing scheme using front detector segments 3--8 (excluding 7). To reconstruct $RHESSI$ images at high energies (50--100 keV and 100--200 keV), we also include front detector segments 1 and 2 to precisely understand the source structure. The sequence of images reveal that, except during peak II, the HXR sources present a compact morphology. It is important to note that the flare began with very high energy HXR emission (see 100--200 keV source in Figure~\ref{X-ray_mosaic}(b)) and this epoch is marked as peak I in the HXR and MW light curves. At this time, although the HXR emitting region is very concentrated, we note two emission centroids in the 50--100 keV image within the compact volume (see also Figure \ref{X-ray_MW_images}(a)). During peak II, the 50--100 keV HXR source exhibits significant morphological evolution with the appearance of a second HXR source at the northern side of the pre-existing source (Figure~\ref{X-ray_mosaic}(c)). After peak II, HXR emission up to 50 keV was observed until 05:09 UT (Figures~\ref{X-ray_mosaic}(e) and (f)). 

In Figure~\ref{X-ray_MW_images}, we plot MW sources at 17 GHz (gray) and 34 GHz (black) over co-temporal HXR images. It is evident that there is a single MW source at both frequencies throughout the flare evolution. We also find that the structure of 34 GHz source resembles the high energy HXR source (i.e., 50--100 keV) except during peak II when the HXR source displayed emission from an extended region with two emission centroids. 

\subsection{UV, EUV and H$\alpha$ Observations}
\label{sub:EUV}

\begin{figure}
\epsscale{1.0}
\plotone{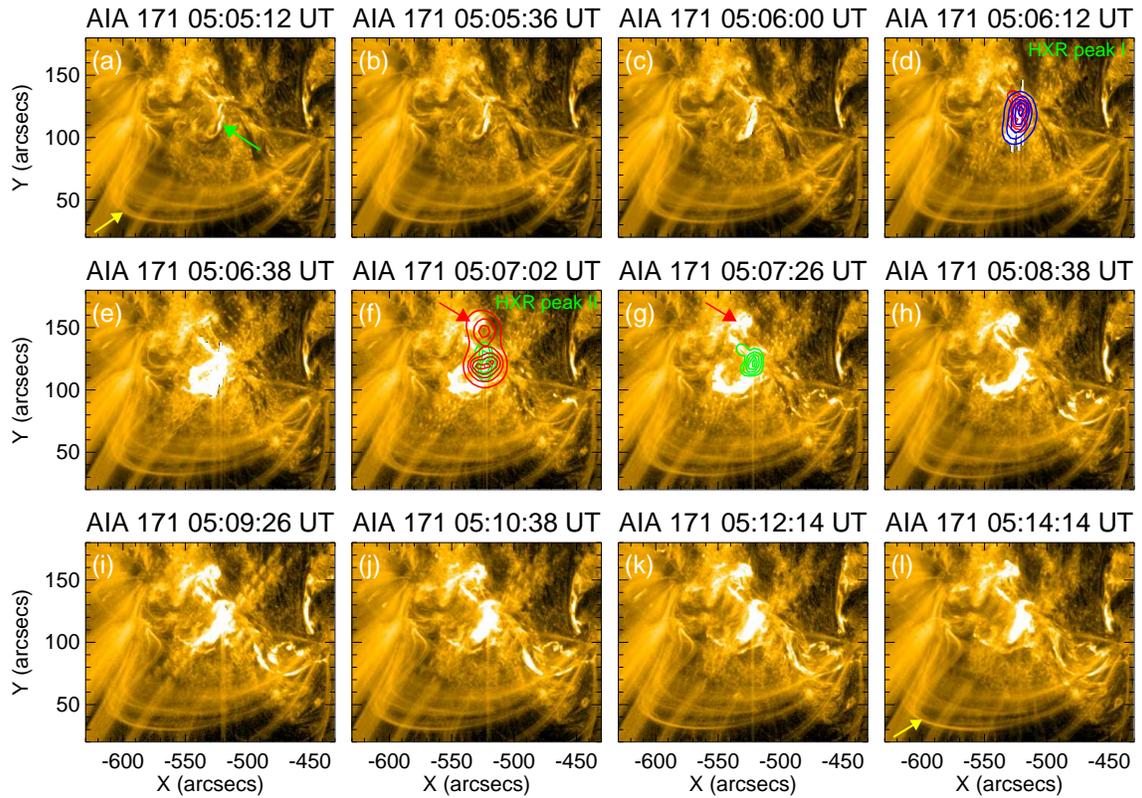}
\caption{Sequence of AIA 171 \AA~ images presenting flare evolution as well as it's magnetic loop environment of the flaring active region. A small J-shaped filament/flux rope is indicated by an arrow (green) in panel (a). Flaring site is fully enveloped by overlying loop systems, which are preserved throughout the event and marked by arrows (yellow) in panels (a) and (l). Red arrows in panels (f) and (g) indicate EUV brightening at a relatively distant location. The $RHESSI$ PIXON images at different energy bands (i.e., 25--50 (green), 50--100 (red) and 100--200 keV (blue) are overplotted on some of the co-temporal 171 \AA~ images. The contour levels for X-ray images are set as 15 $ \% $, 30$\%$, 50$\% $, 70$\%$, and 90$ \% $ of peak flux for each image.}
\label{aia_171}
\end{figure}

In Figure \ref{aia_171}, we present a sequence of AIA 171 \AA~(Fe {\small{IX}}; log($T$)=5.7) images showing important stages of the flare evolution starting from pre-flare stages. The emission at this wavelength originated in a quiet corona and upper transition region. The X-ray images in different energy bands, i.e., 25--50 keV (green), 50--100 keV (red), and 100--200 keV (blue) corresponding to peak I, peak II, and the decay phase are also overplotted in a few selected panels for a comparison. In the very beginning, we note a small J-shaped absorbing structure (filament) in the low corona, which is marked by an arrow (Figure \ref{aia_171}(a)). The flare brightening occurs along this filament which rapidly increases (see Figures \ref{aia_171}(b)--(e)). This early flare location is associated with strong non-thermal HXR emission during peak I (Figure~\ref{aia_171}(d); see also Figures~\ref{X-ray_mosaic}(b)). The flaring region has displayed a rapid increase in the brightness and volume after peak I. During peak II, we observe intense EUV brightening at a new location (red arrow in Figure~\ref{aia_171}(f) and (g)) which is spatially correlated with the second HXR source (Figures~\ref{aia_171}(f) and \ref{X-ray_mosaic}(c)). Following peak II, flare brightening spreads in an extended region and the flare appears as an S-shaped structure (e.g., Figure~\ref{aia_171}(h)). However, we note that there is no signature of the eruption from the flaring region despite intense thermal and non-thermal emission, and the overlying loop structure remains intact (note constant structure of overlying loops which are marked by arrows (yellow) in Figures~\ref{aia_171}(a) and (l)).

\begin{figure}
\epsscale{0.95}
\plotone{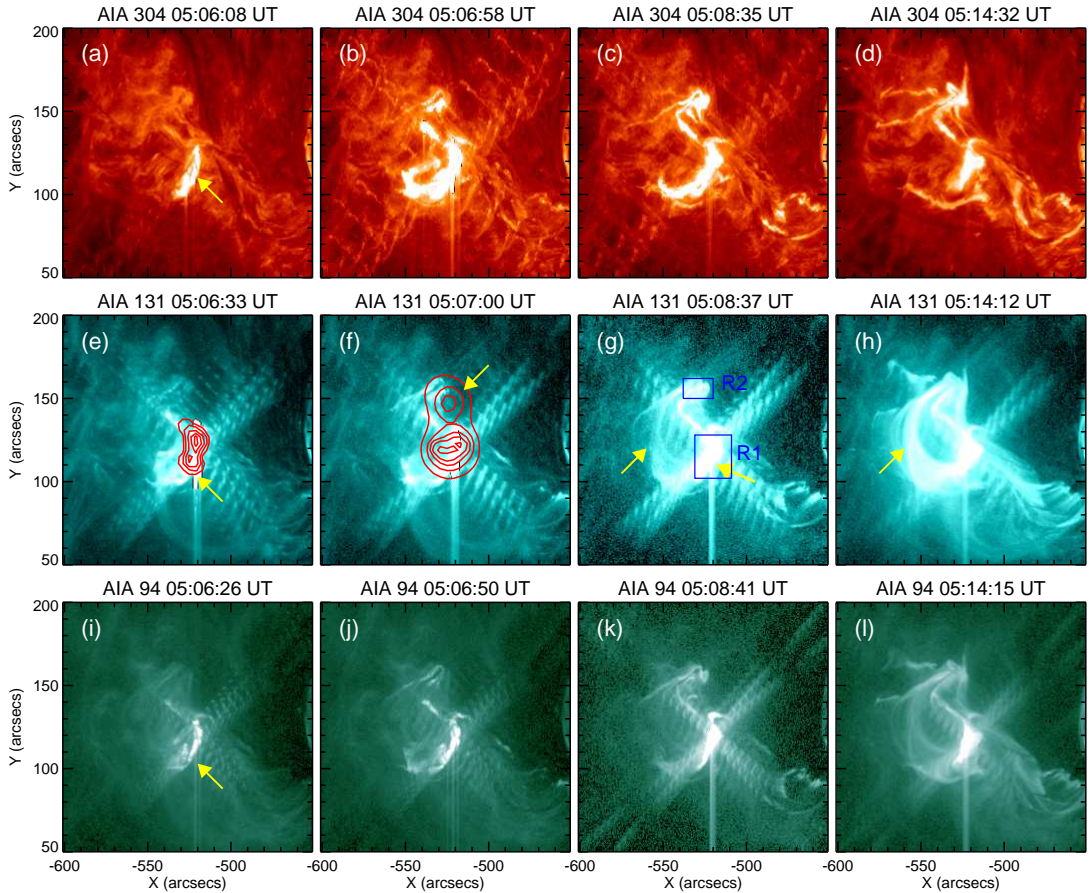}
\caption{Multi-wavelength view of flaring region observed by $SDO$/AIA and $RHESSI$. The contours of $RHESSI$ PIXON images at 50--100 keV (red) energy band are overplotted on some of the representative AIA images (panels (e) and (f)). The position of J-shaped flux rope and its underneath early flare brightening are indicated by arrows (panels (a) and (i)). A number of bright regions are seen at later stages (see panels (c) and (d)). A series of AIA 131 and 94 \AA~images are shown in middle and bottom rows respectively. Hot flaring loop systems are marked by arrows in panels (g) and (h), while their associated two distant footpoint locations are indicated by boxes in panel (g). The contour levels for the X-rays are set as 15$\%$, 30$\% $, 50$\% $, 70 $\%$ and 90$ \% $ of peak flux for each image.}
\label{aia_main}
\end{figure}

In Figure \ref{aia_main}, we present a series of images taken at three different EUV channels (i.e., 304, 131, and 94 \AA) of AIA. The time sequence of the AIA 304 \AA~(He{\small{II}}; log($T$)= 4.7) images provide signatures of the flare evolution in the chromosphere and transition region (top row of Figure \ref{aia_main}). We note a rapid increase in the intensity of the flaring region following peak I, similar to the flare signatures in 171~\AA. In the later stages, we observe several ribbon-like brightenings in the extended region (Figures \ref{aia_main}(c) and (d)). In the middle and bottom rows of Figure \ref{aia_main}, we present sequence of AIA 131 \AA~(Fe {\small{VIII, XXI}}; log($T$)=5.6, 7.0) and AIA 94 \AA~(Fe {\small{XVIII}}; log($T$)=6.8) images showing the transition region and hot flaring corona. The images at these wavelengths provide crucial information on the connectivity of flaring loops and their complex structures. We note an impulsive brightening in short, low-lying loops (marked in Figures~\ref{aia_main}(e) and (i) by arrows) which corresponds to peak I. We call this region as the {\it inner core region}. Following peak II, we observe the formation of a new loop system (marked by arrows in Figures~\ref{aia_main}(g)--(h)) that connects the early flare location (marked as R1; see Figure~\ref{aia_main}(g)) with a relatively remote region (marked as R2; see Figure~\ref{aia_main}(g)). This newly formed loop system essentially forms an {\it outer core region}. The magnetic structure of these two regions, i.e., R1 and R2, is further discussed in Section \ref{sub:EFR} (see also Figure~\ref{fig_hmi_flux_profile}). The R1 region continued to show intense emission till the decay phase. The EUV images taken at 94 \AA~further confirm this scenario (Figures~\ref{aia_main}(i)--(l)).

\begin{figure}
\epsscale{0.90}
\plotone{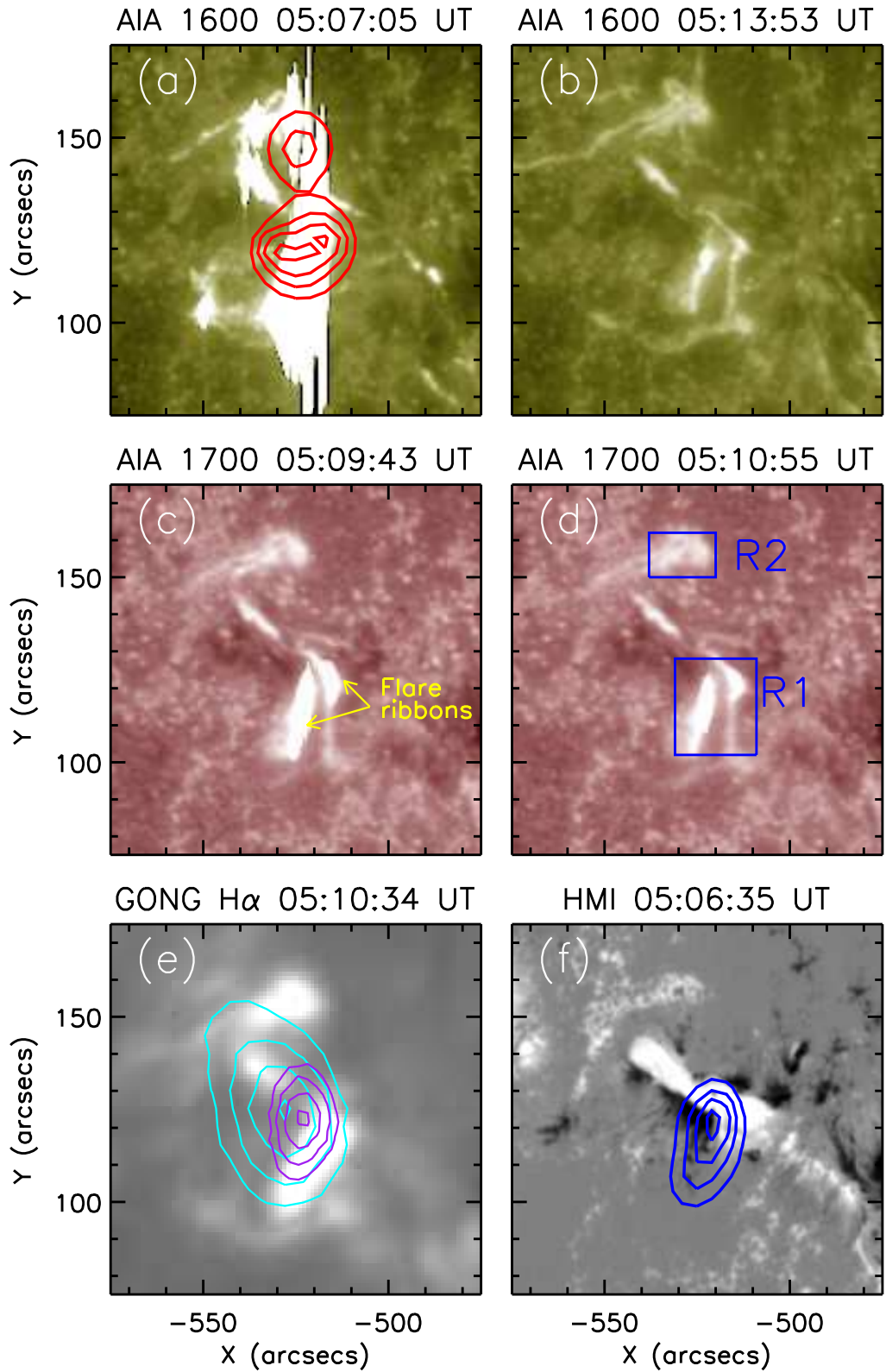}
\caption{Extended view of flaring region in UV and H${\alpha}$. Top and middle rows: AIA observations at 1600 \AA~and 1700 \AA~channels showing the chromospheric and temperature minimum regions of the solar atmosphere. The X-ray contours in 50--100 keV (red) are overplotted on a co-temporal UV image in panel (a). The two distinct parallel bright ribbons at the flaring location are marked as flare ribbons in panel (c). The main flare and remote flare bright locations are enclosed by rectangular boxes R1 and R2, respectively (panel (d)). Panel (e): GONG H${\alpha} $ image overlaid by co-temporal NoRH MW contours at 17 GHz (cyan) and 34 GHz (purple). Panel (f): HMI magnetogram of active region NOAA 11302 overlaid by $RHESSI$ HXR (100--200 keV) source showing the flare location with respect to photospheric magnetic structure.}
\label{aia_decay}
\end{figure}

In Figure \ref{aia_decay}, we present observations taken at 1600~\AA~(C {\small{$IV$}} +cont.; log($T$)=5.0) and 1700~\AA~($continuum$; log($T$)=3.7) channels of AIA along with NSO-GONG H$\alpha $ images during the gradually decaying phase of the flare. For a comparison, $RHESSI$ HXR contours are also overplotted in the AIA 1600 \AA~ image shown in Figure \ref{aia_decay}(a). The observations at these wavelengths present the flare response at lower atmospheric layers such as the temperature minimum photosphere (1700 \AA), chromosphere (H$\alpha$), and upper photosphere and transition region (1600 \AA). In UV images, we can clearly distinguish two well-resolved flare ribbons at the location of the HXR source (Figures \ref{aia_decay}(b)--(d)). Like EUV images, UV observations also show emission from regions R1 and R2 (Figure~\ref{aia_decay}(d)). We also find that H$ \alpha $ and UV features match well, although poor spatial resolution of H$ \alpha $ images do not allow us to resolve H$ \alpha $ flare ribbons. In Figure~\ref{aia_decay}(e), we plot MW 17 and 34 GHz sources over H$\alpha$ images. We find that the 17 GHz emission seems to be associated with hot flaring loops connecting regions R1 and R2. On the other hand, the 34 GHz source is mainly located within R1 region. In Figure \ref{aia_decay}(f), we present an HMI magnetogram of active region NOAA 11302 overlaid by $RHESSI$ HXR (100--200 keV) source showing the flare location with respect to photospheric magnetic structure.

\subsection{RHESSI X-ray Spectroscopy}

We have studied the evolution of $RHESSI$ X-ray spectra during the flare. For this analysis we first generated a $RHESSI$ spectrogram with an energy binning of 1/3 keV from 6--15 keV, 1 keV from 15--100 keV and 3 keV from 100--200 keV energies. We only used the front segments of the detectors and excluded detectors 2 and 7 (which have lower energy resolution and high threshold energies, respectively). The spectra were deconvolved with the full detector response matrix (i.e., off-diagonal elements were included; \citealt{Smith2002})

\begin{figure}
\epsscale{0.85}
\plotone{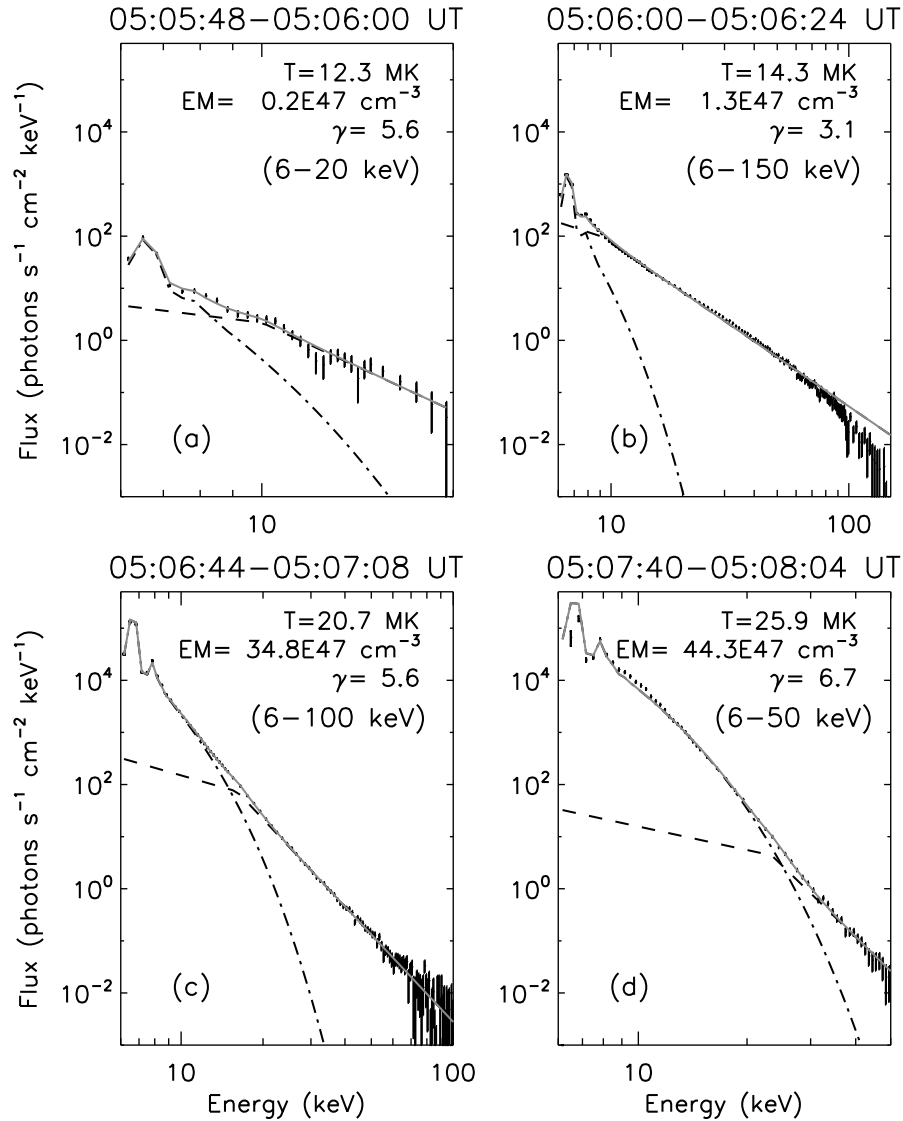}
\caption{$RHESSI$ X-ray spectra derived during the various stages of flare evolution. The spectra were fitted with an isothermal model (dash--dotted line) and a breaking power law with a turnover at low energies (dashed line). The gray (solid) line indicates the sum of the two components. Each spectrum is accumulated over 24 s (except panel (a)) using all $RHESSI$ front detector segments except 2 and 7  and the fitting energy range is indicated in each panel. For the spectra shown in panel (a), we have chosen 12 s integration time because prior to 05:05:48 UT, the non\epsscale{0.80}-thermal HXR emission was insignificant (cf. Figure \ref{rhessi_goes_lc_2}).}
\label{hsi_spectra}
\end{figure}

\begin{figure}
\epsscale{0.85}
\plotone{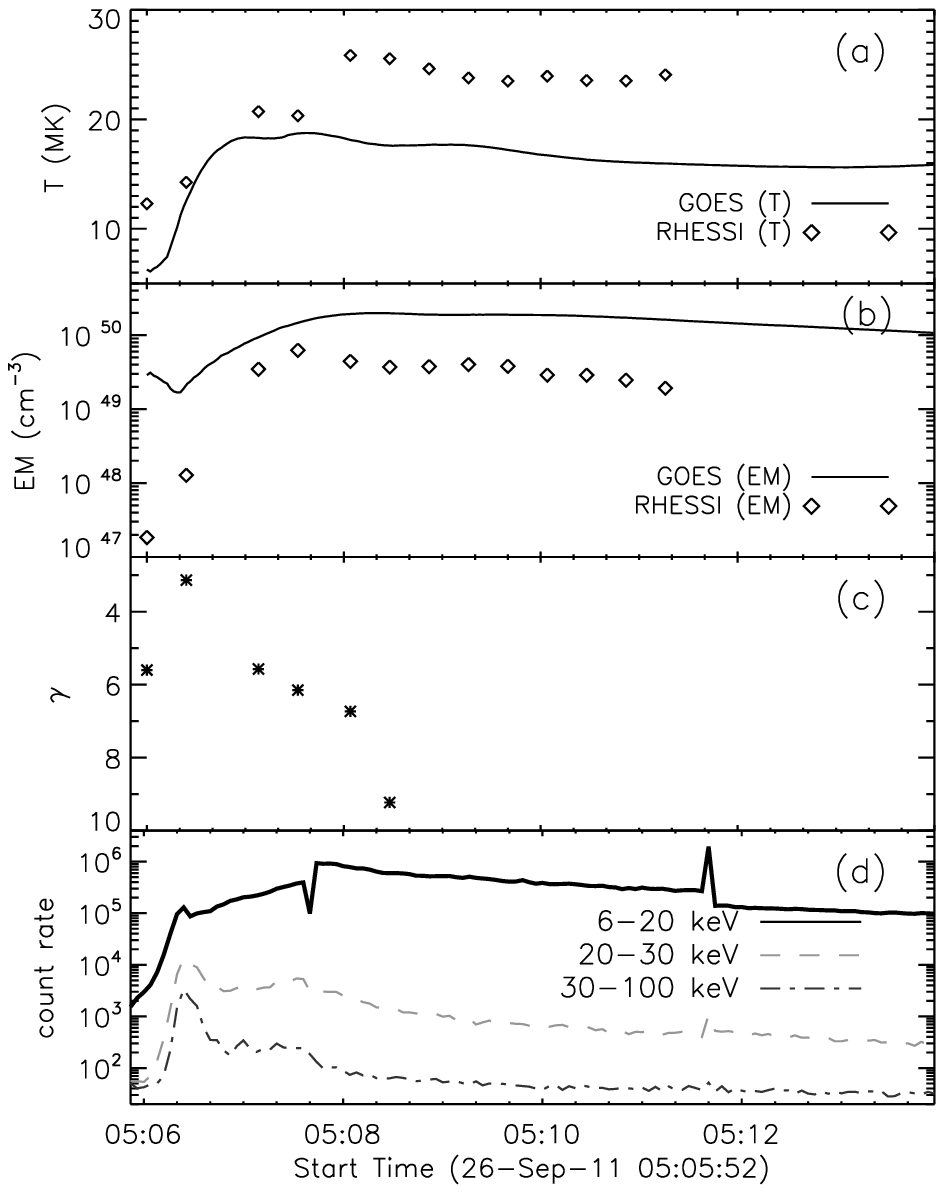}
\caption{Temporal evolution of various spectroscopic quantities derived from $RHESSI$ X-ray spectral fits of consecutive 24 s integration time. From top to bottom: plasma temperature (T), emission measure (EM), photon spectral index ($ \gamma $), and $RHESSI$ count rates in the 6--20 keV, 20--30 keV, and 30--100 keV energy bands. For comparison, we have also shown variations in $T$ and $EM$ obtained from the $GOES$ satellite over the corresponding panels (a) and (b).} \label{spec_param}
\end{figure}

In Figure \ref{hsi_spectra}, we illustrate spatially integrated, background-subtracted $RHESSI$ spectra for four selected intervals. Spectral fits were obtained using a forward-fitting method implemented in the OSPEX code. Since the flare exhibited very fast spectral evolution, the energy range for spectral fittings has been carefully selected for different intervals depending upon the flare and background counts. We used the bremsstrahlung and line spectrum of an isothermal plasma and a power-law function with a turnover at low X-ray energies. The negative power-law index below the low-energy turnover was fixed at 1.5. In this manner, there are five free parameters in the model: temperature ($T$) and emission measure ($EM$) for the thermal component, and power-law index ($ \gamma $ ), normalization of the power law, and low-energy turnover for the non-thermal component. From these fits, we derive the T and EM of the hot flaring plasma as well as $ \gamma $ for the non-thermal component, which are plotted in Figure \ref{spec_param}. We also provide $T$ and $EM$ estimations from the $GOES$ measurements in Figures \ref{spec_param}(a) and (b).

The $RHESSI$ spectra indicate a very impulsive rise of HXR non-thermal emission. There was hardly any X-ray flux above the background before 05:05:40 UT (see Figure \ref{rhessi_goes_lc_2}). As one can clearly recognize, there is weak yet reliable X-ray flux up to 20 keV in the early impulsive phase (see Figure \ref{hsi_spectra} (a)). It is noteworthy that although the HXR level is very low at this interval, the HXR spectrum can be well fitted with a power law index $ \gamma $ = 5.6 at energies $ \sim $10 keV indicating early non-thermal characteristics. The spectra exhibit a very impulsive evolution and indicate strong non-thermal emission starting at very low energies ($ \gtrsim $9 keV) from the next interval onward (Figures \ref{hsi_spectra} and \ref{spec_param}). It is quite evident from Figure \ref{hsi_spectra}(b) that the HXR spectrum follows a hard power law with a photon spectral index $ \gamma $ = 3.1 at the first HXR burst. At this time, HXR emission up to $ \sim $150 keV was observed above the background level (Figure \ref{hsi_spectra} (b)). After the two HXR bursts, the steepness of power law distribution increased and contributions of thermal emission rapidly enhanced up to $ \sim $25 keV. The non-thermal HXR emission was not very significant after $ \sim $05:10 UT. The temperature profile indicates that the temperature was lowest around the first HXR peak (see Figures \ref{spec_param}(a) and \ref{hsi_spectra}(b)). It increases after the HXR impulsive phase and peaked at $ \sim $05:08 UT with $T$~$\sim$26 MK. The plots of $T$ and $EM$ show a good consistency. We further note that the temperature calculated from the two $GOES$ channels are lower than $RHESSI$ measurements (e.g., $GOES$ peak temperature is $\sim$18 MK) but both shows similar trends. The differences between $GOES$ and $RHESSI$ estimations are likely due to the different sensitivity and response of the two instruments. However, this also points to the fact that the observed plasma is multi-thermal \citep[see, e.g.,][]{Li2005}.

\subsection{Structure and Evolution of Magnetic Fields}

\subsubsection{Magnetic Complexity and Flux Emergence}
\label{sub:EFR}
\begin{figure}
\epsscale{0.80}
\plotone{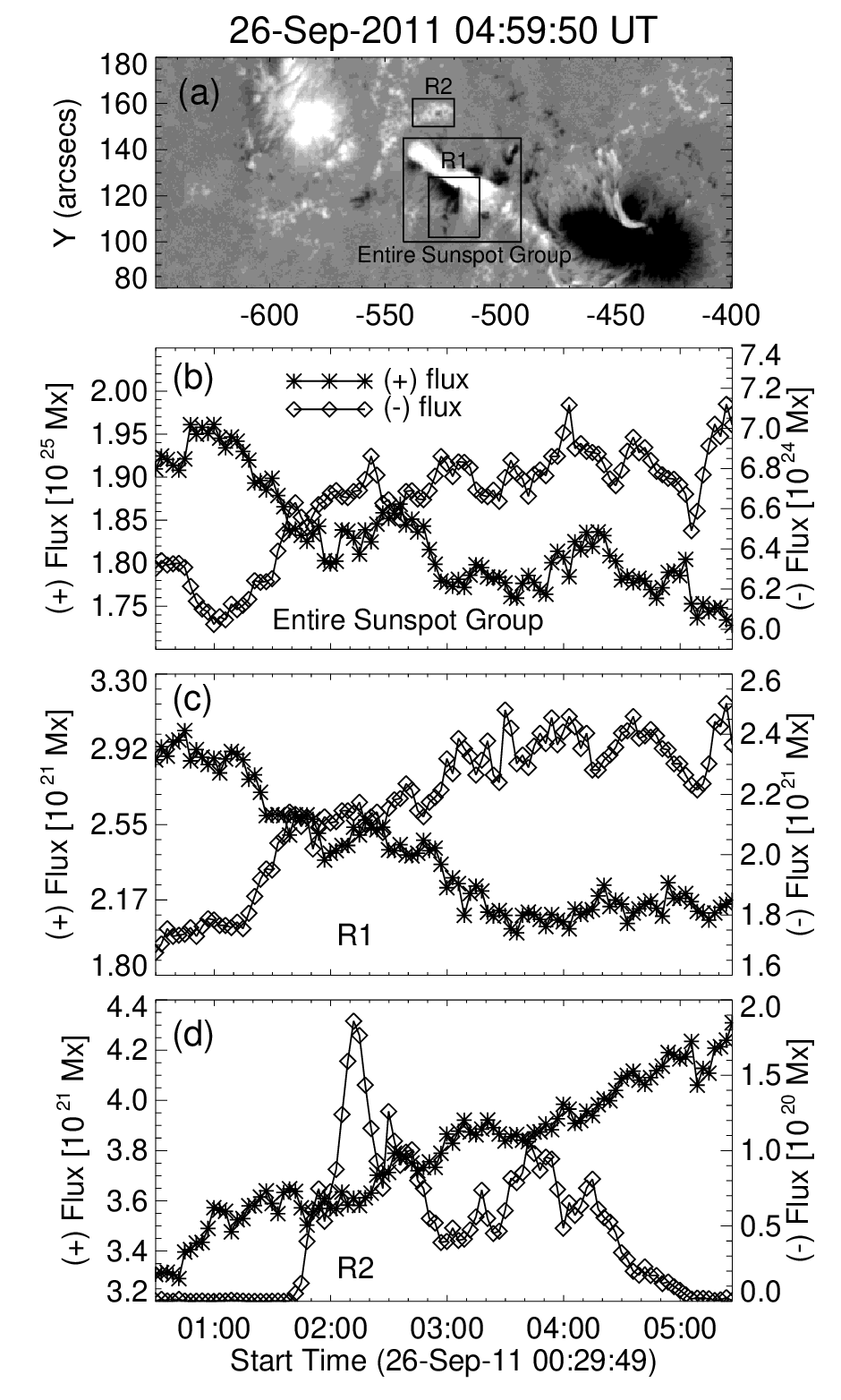}
\caption{Evolution of magnetic flux through different regions associated with flare site. These regions (i.e., entire sunspot group, R1, R2) are shown in panel (a). The magnetic flux of both positive and negative polarities have been computed through these regions which are plotted in panels (b)--(d).}
\label{fig_hmi_flux_profile}
\end{figure}

The active region NOAA 11302 exhibited a very interesting elongated structure consisting of three distinct sunspot groups (see Section~\ref{overview}) and the reported event is associated with the central part of the active region having a mix magnetic polarities (see Figures \ref{event_overview}(a) and (b)). In Figure~\ref{fig_hmi_flux_profile}, we present the time evolution of photospheric magnetic flux derived from $SDO$/HMI magnetograms through different areas of the active region. We use 3-minute averaged data from HMI  between 00:30 UT and 05:30 UT to estimate the photospheric magnetic flux. In order to correct for the solar differential rotation, we have derotated all of the images to a common reference time of 05:30 UT by using the SolarSoftWare routine drot$\_$map.pro. The areas selected for the magnetic flux estimation include the entire sunspot group along with two smaller regions, R1 and R2, which are indicated in Figure~\ref{fig_hmi_flux_profile}(a). We have previously discussed the spatial importance of regions R1 and R2 in (E)UV images (see Figures~\ref{aia_main} and \ref{aia_decay}). R1 essentially corresponds to the early flare location which was also associated with the HXR emission during peak I. In our previous description, we have defined this region as the inner core region (see Section~\ref{sub:EUV}). Region R2 corresponds to the footpoint of a large flare loop system that connects the inner core region to a relatively remote location. 

From the magnetic flux profile of the entire sunspot group (Figure~\ref{fig_hmi_flux_profile}(b)), we find that this region is dominated by the positive polarity. However, the positive flux through this region was observed to decrease continuously while the negative flux exhibits an increasing trend. In the inner core region (R1), we observe a continuous increase of negative flux and consequently positive flux decreases (Figure~\ref{fig_hmi_flux_profile}(c)). We find a very interesting variation of magnetic flux through the R2 region (Figure~\ref{fig_hmi_flux_profile}(d)). It is quite important to note that R2 lies in a magnetically dispersed region of predominant positive polarity, away from the central sunspot group (Figure \ref{fig_hmi_flux_profile}(a)). In this region, we find a continuous increase of positive magnetic flux. On the other hand, the negative flux profile exhibits irregular patterns in its evolution with relatively fast increase and decrease of the magnetic flux. We further note significant fluctuations in the evolution of negative flux between 2:00 and 2:30 UT. A careful examination of the magnetograms reveal that these fluctuations are caused by the appearance of moving magnetic features. It is important to note that the emergence of positive flux through the R2 region and negative flux through the R1 region are of similar scales, suggesting that the two regions are magnetically connected. EUV images after peak II indeed show the existence of bright coronal loops that join the R1 and R2 regions (AIA 131~\AA~and 94~\AA~images in Figures~\ref{aia_main}(g), (h), and (l)).

\begin {figure}
\epsscale{0.90}
\vspace*{-1.75inch}
\plotone{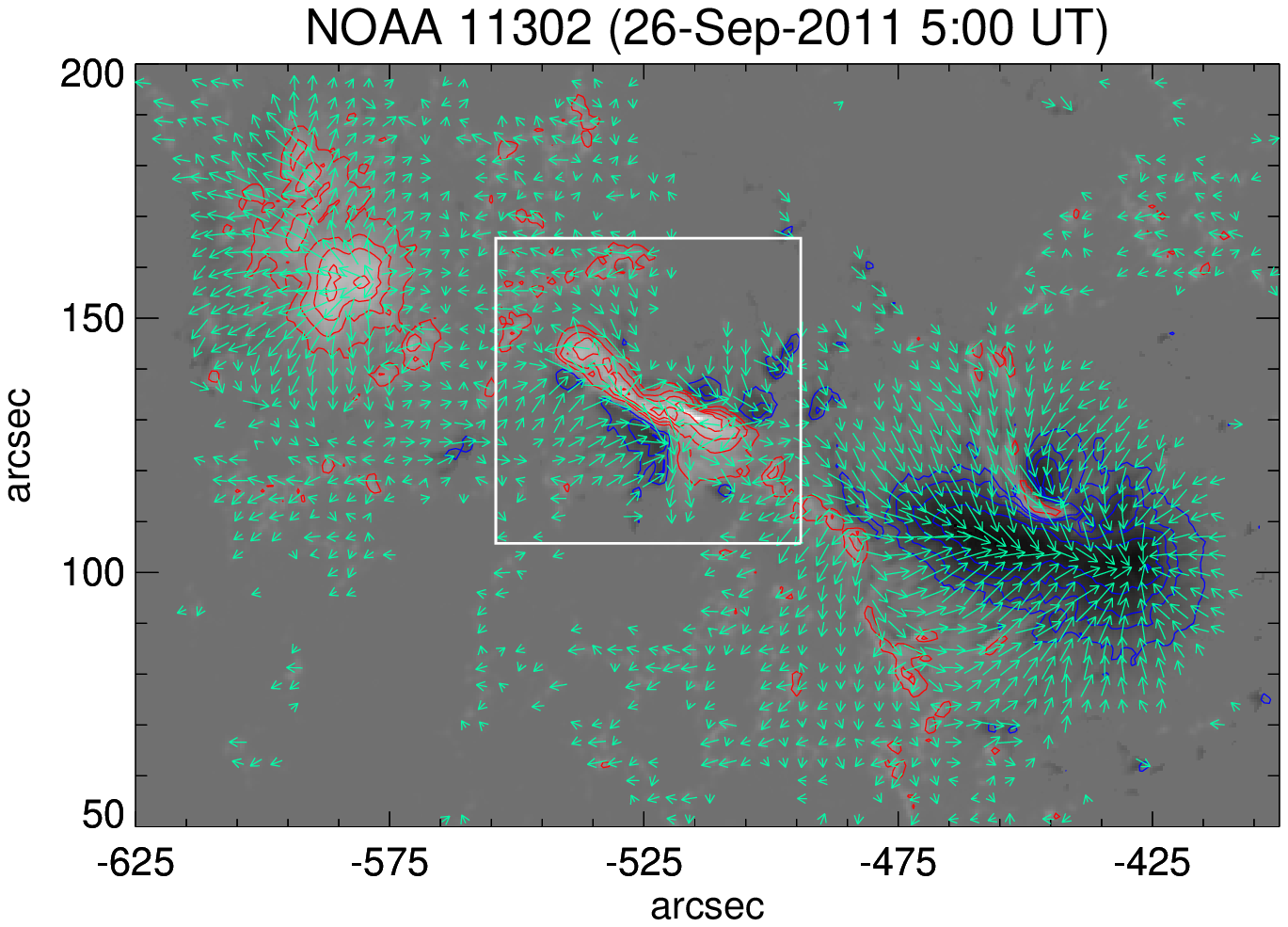}
\caption {Vector map of NOAA AR 11302 obtained from HMI at 05:00 UT on 2011 September 26. Transverse vectors above threshold values are shown as green arrows. Red and blue contours are for positive and negative flux regions with the levels of $\pm$ 2000, $ \pm $ 1500, $ \pm $1000 and $ \pm $500 Gauss. The box shows the region of interest where the flare took place.}  \label{vector2}
\end {figure}

In Figure~\ref{vector2}, we show a vector magnetogram of the active region NOAA 11302 on 2011 September 26 at 05:00 UT obtained from SDO/HMI. The transverse vector fields are indicated by green arrows in the figure. The region associated with flaring activity is shown inside a white box. The direction of transverse vectors between the negative polarity regions and the positive polarity region  indicate that the magnetic field lines are highly sheared. To quantify the non-potentiality of the field lines in the activity site, we compute the spatially averaged signed shear angle \citep[SASSA;][]{Tiwari2009}, which represents the average deviation of the observed transverse vectors from that of the potential transverse vectors. Over the region of interest (shown inside the box in Figure~\ref{vector2}), the SASSA has been estimated as -8$^{\circ}$.82 .

\subsubsection{Emergence of Transient Magnetic Bipole (TMB)}

\begin{figure}
\epsscale{0.95}
\plotone{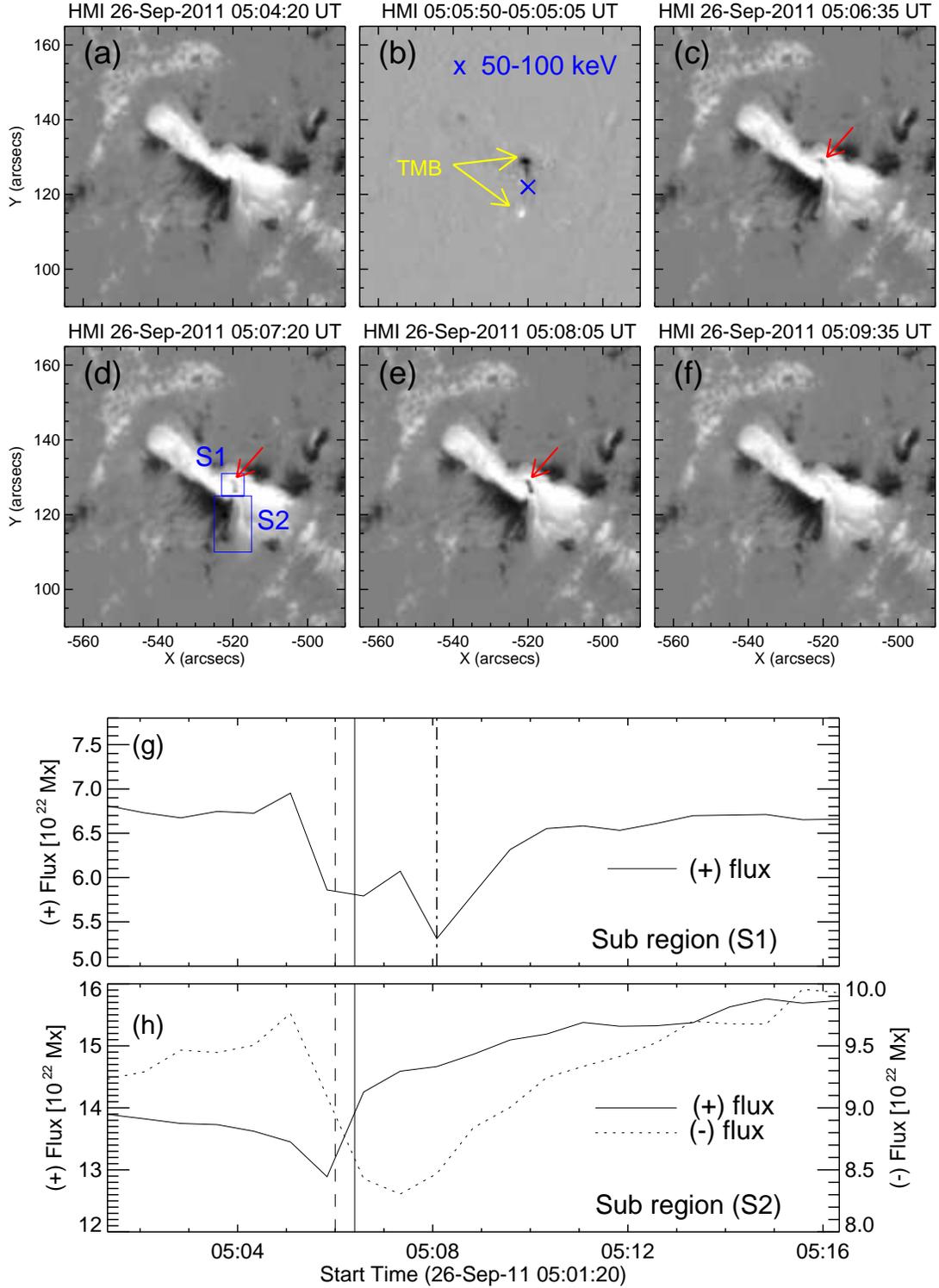}
\caption{Sequence of HMI magnetograms of the flaring sunspot group, showing magnetic field evolution during the event. Panel (b) presents a difference image of two consecutive magnetograms of just before the flare onset time. The emergence of transient magnetic bipole (TMB) is indicated by a V-shaped arrow in panel (b). The ``$\times $" symbol marks the centroid of the X-ray source in 50--100 keV (blue). An intrusion of negative polarity flux into the positive polarity region is indicated by red arrow in panels (c)--(e). The two sub-regions of different sizes are selected for the quantitative study of flux evolution which are shown by rectangular boxes in panel (d). In the panels (g) and (h), we plot the flux profiles of sub-regions S1 and S2 respectively. Dashed and solid vertical lines mark the flare onset and peak of the impulsive phase. Dash-dotted line in panel (g) indicates the time when the positive flux cancellation through the sub-region S1 was maximum.}
\label{fig_hmi_bmts}
\end{figure}

A careful examination of HMI magnetograms reveals the emergence of a spatially separated bipolar magnetic region within the inner core region which developed about $\sim$1 minute prior to the flare onset (Figure~\ref{fig_hmi_bmts}). In Figure~\ref{fig_hmi_bmts}(b), we show a running difference image of magnetograms taken at the pre-flare phase. This difference image clearly displays the emergence of a magnetic bipole (shown by arrows). Here it is worth emphasizing that this transient magnetic bipole (TMB) appeared just before the flare onset and cannot be seen in earlier magnetograms (which are taken at a time cadence of 45~s). This implies that the emerging bipolar region evolved rather quickly and was separated, presumably due to a pre-existing highly sheared magnetic structure.  We further note that the negative polarity region appears as an intrusion of magnetic flux in a larger region dominated by positive polarity (marked by arrows in Figures~\ref{fig_hmi_bmts}(c)--(e)). For a qualitative analysis of the magnetic flux evolution, we have enclosed the negative flux region within a box S1 as shown in Figure~\ref{fig_hmi_bmts}(d). We have measured the positive and negative flux through the sub-region S1 for all the available magnetograms from 05:01 to 05:16 UT observed with a time cadence of 45 s. We find that with the appearance of the intrusion of the negative flux region at 05:05:05 UT, the positive flux through sub-region S1 started to decrease when the maximum cancellation of positive flux occurred at 05:08:05 UT (see Figure \ref{fig_hmi_bmts}(g)). Here, it is important to note that the negative flux through S1 was nil during this interval except at 05:08:05 UT when the total negative flux through this region was 6.4 $ \times $10$ ^{18} $ Mx. Although the cause of the magnetic transient is unknown, we can speculate that the newly emerged negative flux was initially entirely depleted in canceling the pre-existing strong, large-scale positive flux. We note that the flux started decaying from 05:05:05 UT, i.e., at about 1 minute prior to the impulsive flare emission. The onset of the flux decay with respect to the triggering of flare can clearly be seen in Figure \ref{rhessi_goes_lc_2}, where the dashed vertical line marks the time at which flux cancellation begins.     

HMI magnetograms clearly indicate that the positive flux region has a relatively extended structure which is shown inside the rectangular box S2 (Figure \ref{fig_hmi_bmts}(d)). The time profiles of magnetic flux evolution through S2 are plotted in Figure \ref{fig_hmi_bmts}(h), which clearly indicates an increase in positive flux (and a corresponding decrease in negative flux) throughout the flare evolution. We also find that the main HXR source is localized in between the TMB (Figure~\ref{fig_hmi_bmts}(b)).  

\subsubsection{Spectral Behavior of TMBs }
 
The emergence of TMB prior to the event is a very important phenomenon as it occurs in a localized region over relatively short timescales. Therefore, it is desirable to probe the characteristics of these small-scale magnetic structures more rigorously to provide a plausible interpretation of this phenomena in relation to the flare occurrence. 

\begin{figure}

\epsscale{0.95}
\plotone{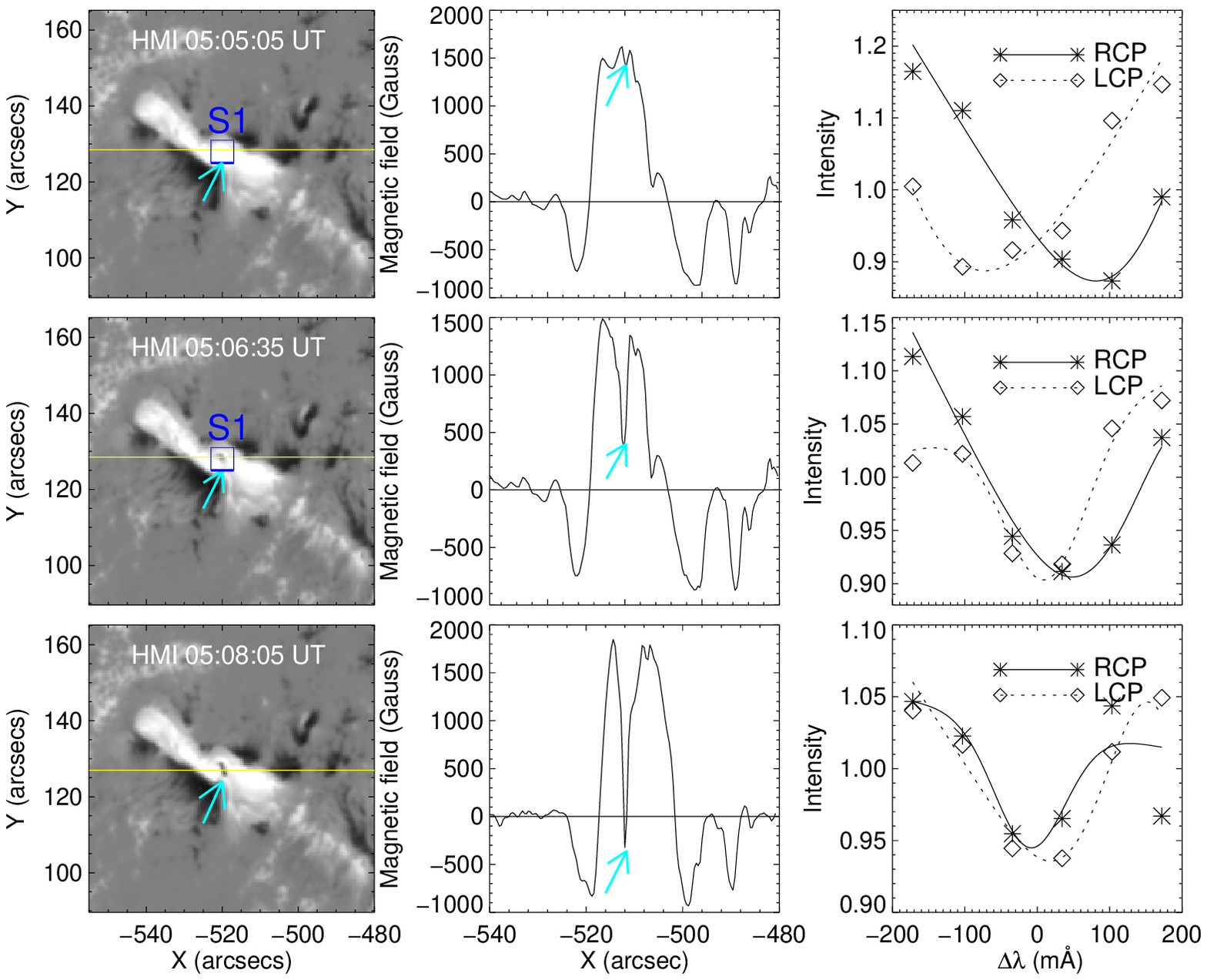}
\caption{Left column: sequence of HMI magnetograms. Middle column: magnetic flux profiles along a horizontal line passing through sub region S1. Right column: HMI six-point spectra in both LCP and RCP components of Fe {\small{I}} line for the respective locations marked by arrows (cyan) in the left column.}
\label{fig_spectra1}
\end{figure}

\begin{figure}
\plotone{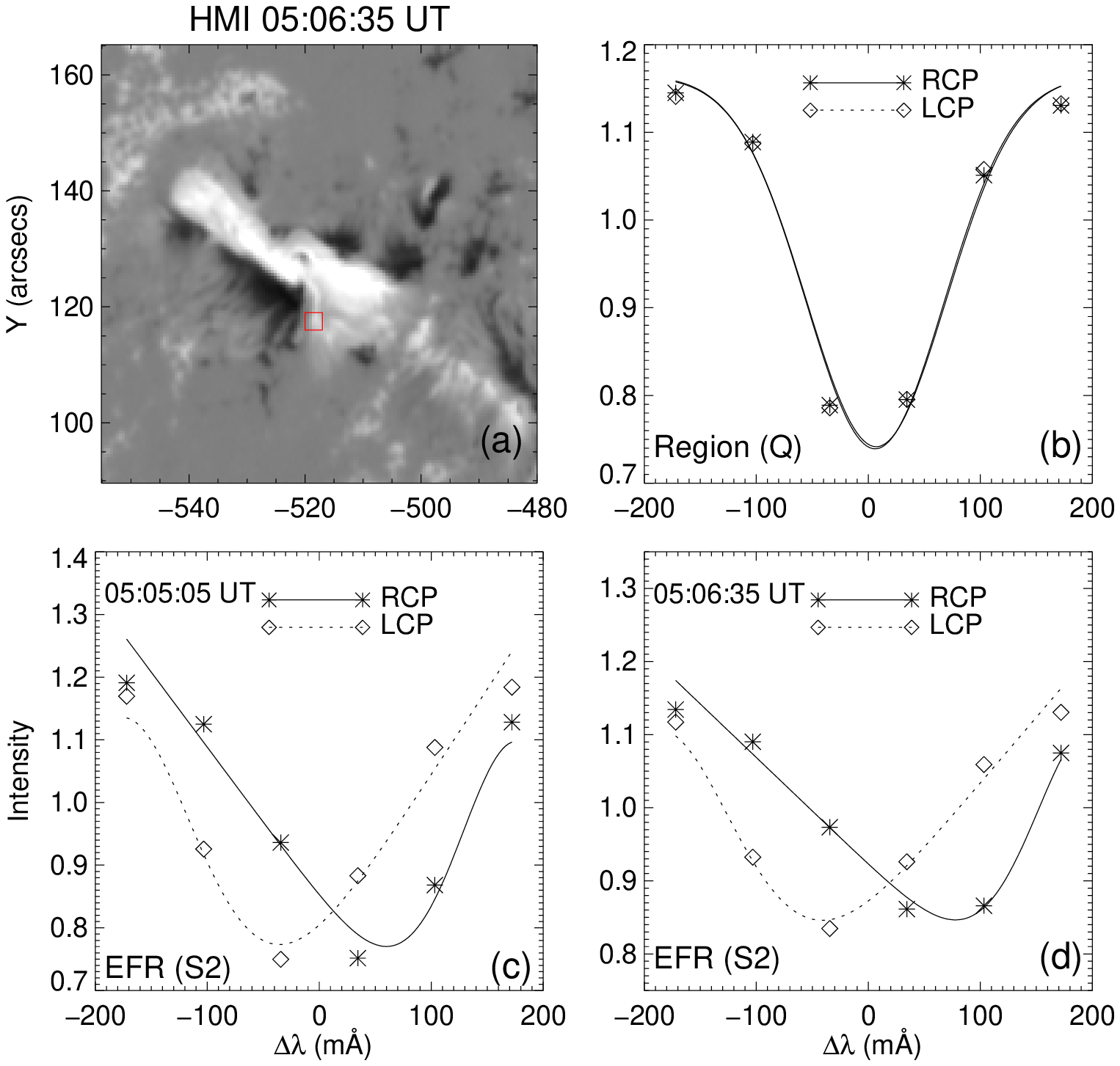}
\caption{Top left: HMI magnetogram of activity site around the peak time of the event. The region selected for the spectral analysis is shown in panel (a) by a small box. Top right: LCP and RCP profiles for a quiet region (nil magnetic field). Bottom left: HMI six-point spectra for (+) EFR at the flare onset time (05:05:05 UT) in both LCP and RCP components. The splitting between the centroid of the LCP and RCP profiles is $ \sim $70 m\AA~ and the RCP profile shifted toward positive values. Bottom right:  spectral profiles of same region at the peak time of the event (05:06:35 UT). At this time the splitting between the centroid of LCP and RCP is $\sim $90 m\AA.} \label{fig_spectra2}
\end{figure}

In Figure~\ref{fig_spectra1}, we investigate the variations of magnetic flux in the negative emerging flux region. In the left column of Figure~\ref{fig_spectra1}, we show a sequence of three representative HMI magnetograms in which a horizontal line is drawn that passes through the negative flux region. The magnetic flux profiles along these horizontal lines are plotted in the middle column of Figure~\ref{fig_spectra1}. A comparison of these magnetic flux profiles clearly indicates a continuous cancellation of positive magnetic flux at the location of the negative polarity transient (the location of the transient is marked by arrows in right and middle columns of Figure~\ref{fig_spectra1}). The plots in the right column of Figure~\ref{fig_spectra1} show HMI spectral profiles at the location of the magnetic transient. The spectral profiles show the variation of intensity at six wavelengths ($\pm$34.4, $\pm$103.3, and $\pm$172.0 m\AA) around the center of Fe {\small{I}} line at 6173.3 \AA~for left circular polarization (LCP) and right circular polarization (RCP). For reference, line profiles corresponding to a magnetically quiet region are shown in Figure \ref{fig_spectra2}(b), which obviously illustrate an overlap of LCP and RCP profiles. It is evident that with the cancellation (decrease) of positive flux, the two line profiles (i.e., LCP and RCP) shift systematically toward each other and eventually switch positions with the reversal of magnetic polarity (from the top to bottom panels of Figure \ref{fig_spectra1}). Similarly, the HMI spectral line profiles corresponding to the region of emerging positive flux region is presented in Figure \ref{fig_spectra2}.

\section{Discussion and Conclusion}
\label{sec:res}

We present a detailed multi-wavelength study of an M4.0 flare that occurred in NOAA AR 11302 on 2011 September 26. Table \ref{tab1} presents observational summary of various multi-wavelength phenomena associated with this event. The active region NOAA 11302 had a peculiar elongated structure with three separated  sunspot groups which altogether formed a $ \beta\gamma\delta $ magnetic configuration. The M4.0 flare is associated with the central sunspot group. From HMI magnetograms, it is quite evident that this central region exhibited mixed distribution of polarities in the photosphere. 

We note that the present event is an appropriate example of confined flare in that it was a short-lived event that occurred in the low corona and did not lead to a CME \citep[see e.g.,][]{Wang2007}. The absence of the ejection of any material from the flaring site is further evident from the fact that the overlying large magnetic loops, imaged in AIA 171~\AA, remain unaltered following the flare. On the other hand, the flare is accompanied with very high energy HXR emission (up to 200 keV) and at the peak we detected an HXR source at 100--200 keV, evidencing an intense energy release in the form of non-thermal particles. Therefore, the absence of an eruption from such a high energetic flare implies that the core fields were not allowed to escape by the overlying fields despite the intense energy release. \cite{Wang2007} have made a systematic study of a set of X-class flares of eruptive and confined categories. Their results suggest that confined events occur closer to the magnetic center, while eruptive events tend to occur close to the edge of active regions. For a flare occurring at the central part of an active region, like the present one, it is more likely that the core field region will have a stronger envelope of overlying fields which may act as a shield for the prevention of eruption \citep[][]{Torok2004,Torok2005}. 

The temporal evolution of HXR and MW emissions provide crucial evidence of the energy release processes. Both $RHESSI$ and NoRH light curves reveal very abrupt rise of flare emission that peaked instantly and simultaneously in both HXR and MW channels. The increase in HXR flux at energies $>$25 keV is usually identified as the start of the impulsive phase of a flare which is suggestive of the injection of the non-thermal electrons into the flaring loop \citep[see, e.g.,][]{Siarkowski2009,Joshi2012}. In our observations we do not find any pre-cursor or pre-flare enhancements in X-ray profiles before the impulsive phase. We further emphasize that the $GOES$ low-energy channels also indicate an increase in X-ray flux with the beginning of the HXR impulsive phase. The rapid temporal evolution of X-ray and MW flux along with strong non-thermal characteristics of X-ray spectra ($ \gamma\sim$3) indicate that this event is an ``early impulsive flare" \citep{Sui2006}. We further stress that at the peak of the impulsive phase the transition from thermal to non-thermal emissions took place at a very low energy ($ \sim $9 keV) while the spectrum extended up to 150 keV. The lack of precursor phase \citep{Joshi2011} clearly implies the absence of significant plasma pre-heating before strong particle acceleration. Due to the confined nature (i.e., non-eruptiveness) of this highly impulsive event, we interpret that the early non-thermal HXR and MW bursts are the consequence of abrupt energy release via spontaneous magnetic reconnection \citep[see, e.g.,][]{Lee2003}.

The impulsive phase lasted for $ < $1 minute. It is important to note that, within this short phase, the HXR and MW time profiles displayed dual peaks at a separation of $ \sim $32 s. It should be noted that these two peaks differ in terms of spectral properties of the HXR emission as well as the morphology of the HXR source structure. (1) The thermal emission rapidly built up after the first HXR peak while non-thermal contribution significantly decreased (e.g., turnover from thermal to non-thermal energies increased from $ \sim $9 keV to $ \sim $15 keV). (2) At the first peak, the HXR emitting region was highly concentrated with multiple centroids and was observed up to 100--200 keV energies while the second HXR peak was relatively less energetic and a new 50--100 keV HXR source originated for a very brief interval that was well separated from the pre-existing HXR source location. These observations lead us to believe that the dual-peaked structure of the impulsive phase corresponds to the distinct events of electron acceleration in the corona \citep{Grigis2004,Kundu2009,Li2005} as a result of a two-step magnetic reconnection process.       

The combined multi-wavelength view of the flare imaged by $SDO$/AIA in several EUV and UV wavelengths provides a clear scenario of coronal energy release and its response at lower atmospheric layers.  The observations taken in all the EUV channels reveal that the flare was initiated with a sudden brightening at a very localized region above/around a small filament. The co-spatiality of this localized EUV region (defined as \textit{inner core region} in the previous section) with the high-energy HXR source (e.g., 100--200 keV) provide evidence for the strong particle acceleration at this location. It is worth mentioning that the HXR emission from the inner core region is highly concentrated and lasted untill the decay phase of the flare. We further note that the HXR sources at lower, intermediate, and higher energies are nearly co-spatial and do not indicate significant spatial evolution. In the standard flare model, HXR emission is believed to originate from distinct locations of flaring loops, viz. footpoint and looptop regions \citep{Krucker2008}. In general, HXR emissions at lower energies originated from the coronal region while the high-energy part is associated from denser chromospheric layers where footpoints of the loops are anchored \citep[see, e.g.,][]{Joshi2009,Joshi2007}. In our case, the overlapping of high- and low-energy sources possibly indicates that looptop and footpoint sources of a flaring loop are too close to each other to be resolved, i.e., at least one of the loops involved in the magnetic reconnection is a very short low-lying loop. This is further confirmed by UV images taken at 1600 and 1700 \AA~which show very closely situated flare ribbons formed following the initiation of magnetic reconnection. Closely separated flare ribbons would imply formation of short, simplified loops after the primary energy release. It is likely that due to small separation, conjugate HXR sources corresponding to the two ribbons are not distinguishable. 

Peak II of the impulsive phase is associated with EUV brightening in a relatively remote location besides  continuous emission from the inner core region. It is important to note the high-energy HXR emission (50--100 keV source) that originated from this remote flare location for a very brief interval. More importantly, we observed high-temperature EUV loops (e.g., 131 \AA~and 94 \AA~images) that connect this new source with the inner core region. The configuration and evolution of new loops form an \textit{outer core region} which showed intense thermal emission after the impulsive phase. The formation of new loop system suggests a second stage of magnetic reconnection during which short loops in the inner core region interacted with larger coronal loops at relatively higher altitudes. Thus the sequential brightening of inner and outer regions of coronal field lines in the core region provide evidence for flare models involving the interaction of coronal loops \citep[see e.g.,][]{Hanaoka1996,Hanaoka1997,Kundu2001}. Although it is difficult to reconstruct the geometry of the loops for this event, the configuration of newly developed loop with respect to the inner core region suggests that the two interacting loop systems probably shared a common footpoint. \cite{Hanaoka1997} proposed that in such a double-loop configuration, a flare can be caused by the interaction between an emerging loop and an overlying loop.     

A crucial aspect of this study lies in exploring the phenomenon of flux emergence both at large and relatively smaller spatial and temporal scales. Large-scale changes in photospheric magnetic flux were observed from $ \sim $5 hr prior to the flare. A continuous increase of negative flux was detected at the main flare location (R1; cf. Figure \ref{fig_hmi_flux_profile}), while positive flux emergence occurred at a relatively remote location (R2; cf. Figure \ref{fig_hmi_flux_profile}). Moreover, EUV observations clearly show the formation of a loop system that joins the R1 and R2 regions which provide clear evidence that the two regions are magnetically connected. The analysis of HMI vector magnetograms clearly reveal highly sheared magnetic structures in pre-flare phases with SASSA as 8$^{\circ} $.8. It is likely that the continuous flux emergence would increase the shearing of field lines. Earlier studies have shown that SASSA $ \gtrsim $8$^{\circ} $ indicates that the active region is capable of driving major flares of M and X categories \citep{Tiwari2010}. The flux emergence along with enhanced magnetic shear is believed to be the fundamental process associated with the supply of magnetic free energy into the corona in the pre-flare stages \citep{Wang2004,Zheng2012}. 
 
 The emergence of a pair of magnetic transients of opposite polarity (called TMB in preceding sections) within the inner core region has important physical implications for recognizing the processes that lead to the destabilization of the magnetic structure and thereby causing the flare initiation. The role of small magnetic structures toward the flare triggering processes has been addressed in simulations in a recent study by \cite{Kusano2012}. Since TMB reported here appear about 1 minute prior to the flare onset, relating this feature with the flare triggering mechanism is justified. We stress that out of two magnetic transients, the negative polarity transients is highly distinctive due to its location and morphology (note the S1 region shown in HMI magnetograms in Figure \ref{fig_hmi_bmts}). A patch of negative polarity intruded in a larger region of positive flux and exhibited a compact structure that remained stationary. In a recent study, \cite{Harker2013} reported a very similar magnetic feature which indicates small-scale changes in the magnetic field structure.
 
 The HMI spectra of the TMB clearly indicate the shifting of LCP and RCP profiles in wavelength direction which suggest a systematic dominance of corresponding magnetic polarities from the pre-flare phase (Figures \ref{fig_spectra1} and \ref{fig_spectra2}). Further the spectral profiles do not indicate any flare-induced anomaly, such as sudden deviation of the spectral line from absorption to emission \citep{Maurya2012,Qiu2003}. These observations coupled with the fact that magnetic transients developed before the flare and located away from the HXR flare kernels provide evidence that these magnetic transients represent a real change in the magnetic field structure preceding the flare. 
 
 The confined M4.0 flare on 2011 September 26 analyzed here occurred at the core of the complex active region NOAA 11302. The large-scale flux emergence and high magnetic shear at the photospheric level imply that a large amount of free energy was stored in the overlying coronal loops. Here the most crucial observation is that the flare triggering (and subsequent X-ray emission) occurred just after the emergence of a pair of small-scale magnetic transients of opposite polarity at the inner core region of the flaring environment. The flare underwent a confined eruption and displayed a compact morphology. Furthermore, the flare did not show any pre-flare or precursor thermal emission. The present study suggests that small-scale changes in the magnetic structure play a crucial role in the flare triggering process by disturbing the pre-existing sheared magnetic configuration. The interpretation of such magnetic transients developing prior to the flare onset is ambiguous i.e., we cannot yet determine whether these features are associated with the real emergence of magnetic flux or structural changes in the magnetic configuration. Since magnetic transients reported here developed simultaneously as a pair of opposite polarity patches and the corresponding regions exhibit variations in the magnetic flux from the pre-flare to the end of impulsive phase, we cannot rule out the possibility that these magnetic transients resulted from the impulsive emergence of magnetic flux through the photosphere. To our knowledge, such rapidly evolving bipolar magnetic structures prior to the flare has been reported for the first time. For a better understanding of the role of small-scale magnetic structures toward the flare initiation process, a comprehensive survey of existing and upcoming high-resolution solar data is required. In the future, we plan to probe the characteristics of magnetic transients in detail.   

\begin{table}
\begin{center}
\caption{Observational summary of activities before and during confined M4.0 flare.}
\smallskip

\begin{tabular}{p{1.5in}p{1.20in}p{0.9in}p{2.2in}}
\tableline\tableline
Observation/Activity & Time  & Wavelength & Remarks\\

\tableline
Emergence of magnetic flux & From $\sim$5 hr prior to the flare & Magnetogram & Flux emergence of negative and positive polarities in the inner and outer core regions associated with flare loop configuration respectively\\

Small J-shaped filament at the flare location & Pre-flare images ($<$05:06 UT) & EUV (e.g., 304, 171 \AA) & Filament is seen as an absorbing feature\\  

Appearance of transient magnetic bipole (TMB) & From pre-flare phase (05:05:05 UT) to end of impulsive phase & Magnetogram & $ \sim $1 minute prior to flare onset \\

Confined M4.0 flare in the vicinity of filament & Impulsive phase started at 05:06:00 UT with peak emission at 05:06:24 UT & HXR, MW and EUV & Two distinct non-thermal peaks at interval of 32 s, evidence for a two-step magnetic reconnection process \\ 

Impulsive phase & 05:06--05:08 UT &  HXR ($ >$25 keV)& HXR emission up to 200 keV energies with $\gamma$ $\sim $3, mostly a single, compact HXR source except during peak II\\ 
Evolution of distinct inner and outer core regions of the flare loop system &  05:06--05:10 UT & EUV &  Large overlying loops of the active region remain intact\\
  
\tableline
\label{tab1}
\end{tabular}
\end{center}
\end{table}

\acknowledgments
We thank the $SDO$, $RHESSI$, NoRH, and $GOES$ teams for their open data policy. $SDO$ and $RHESSI$ are NASA's missons under living with a star and small explorer programs, respectively. We are grateful to S. Couvidat of Stanford University for providing the $SDO$/HMI spectral data. We express our sincere gratitude to P. Venkatakrishnan (USO/PRL), Jongchul Chae (Seoul National University), and Yong-Jae Moon (Kyung-Hee University) for useful discussions. KSC is supported by the ``Development of Korea SpaceWeather Center" of KASI and the KASI basic research funds. A.V. gratefully acknowledges the Austrian Science Fund (FWF): P24092-N16. S.K.T. is supported by an appointment to the NASA Postdoctoral Program at the NASA Marshall Space Flight Center, administered by Oak Ridge Associated Universities through a contract with NASA. We sincerely thank the anonymous referee for providing constructive comments and suggestions which have enhanced the quality and presentation of this paper. 



\end{document}